\documentclass[11pt]{article}
\usepackage{fancyhdr}
\usepackage{amsmath}
\usepackage{mathtools}
\usepackage{amsfonts}
\usepackage{amsmath}
\usepackage{amsbsy}
\usepackage{latexsym}
\usepackage[T1]{fontenc}
\usepackage[british,UKenglish,USenglish,english,american]{babel}
\usepackage{graphicx}
\newcommand{\RR}{\mathbb R}
\newcommand{\CC}{\mathbb C}
\newcommand{\NN}{{\mathbb N}}
\newcommand{\ZZ}{{\mathbb Z}}

\newcommand{\sign}{\mbox{sign}}

 \newcommand{\beq}{\begin{equation}}
	\newcommand{\eeq}{\end{equation}}
\newcommand{\ba}{\begin{array}}
	\newcommand{\ea}{\end{array}}
\newcommand{\bea}{\begin{eqnarray}}
	\newcommand{\eea}{\end{eqnarray}}
\newcommand{\eps}{{\epsilon}}

\usepackage{graphicx}
\usepackage{xcolor}

\DeclareMathAlphabet{\mathpzc}{OT1}{pzc}{m}{it}

\begin{document}

\begin{center}
{\bf Quasi one dimensional anomalous (rogue) waves in multidimensional nonlinear Schr\"odinger equations 1:\\ fission and fusion}
\vskip 15pt
{\it F. Coppini $^{1,2,3}$ and P. M. Santini $^{1,4}$}

\vskip 20pt

{\it
$^1$ Dipartimento di Fisica, Universit\`a di Roma "La Sapienza", and \\
Istituto Nazionale di Fisica Nucleare (INFN), Sezione di Roma, \\ 
Piazz.le Aldo Moro 2, I-00185 Roma, Italy\\
$^3$ University at Buffalo, New York, USA. Department of Mathematics.
}

\vskip 10pt
$^{2}$e-mail:  {\tt fcoppini@buffalo.edu, francesco.coppini@uniroma1.it\\ francesco.coppini@roma1.infn.it}\\
$^{4}$e-mail:  {\tt paolomaria.santini@uniroma1.it \\ paolo.santini@roma1.infn.it}\\
\bigskip
\vskip 10pt

{\today}

\end{center}

\begin{abstract}
  In this paper we study the first nonlinear stage of modulation instability (NLSMI) of $x$-periodic anomalous waves (AWs) in multidimensional generalizations of the focusing nonlinear Schrödinger (NLS) equation, like the non-integrable elliptic and hyperbolic NLS equations in $2+1$ and $3+1$ dimensions. In the quasi one-dimensional (Q1D) regime, where the wavelength in the $x$ direction of propagation is significantly smaller than in the transversal directions, the behavior at leading order is universal, independent of the particular model, and described by adiabatic deformations of the Akhmediev breather solution of NLS.  Varying the initial data, the first NLSMI shows various combinations of basic processes, like AW growth from the unstable background, followed by fission in the slowly varying transversal directions, and the inverse process of fusion, followed by AW decay to the background. Fission and fusion are critical processes showing similarities with multidimensional wave breaking, and with phase transitions of second kind and critical exponent $1/2$. In $3+1$ dimensions with radial symmetry in the transversal slowly varying plane, fission consists in the formation of an opening smoke ring centered on the $x$ axis. In the long wave limit, the Q1D Akhmediev breather reduces to the Q1D analogue of the Peregrine instanton, rationally localized in space. Numerical experiments on the hyperbolic NLS equation show that the process of "AW growth + fission" is not restricted to the Q1D regime, extending to a finite area of the modulation instability domain. The universality of these processes suggests their observability in natural phenomena related to AWs in contexts such as water waves, nonlinear optics, plasma physics, and Bose-Einstein condensates.

\end{abstract}

\section{Introduction}

The self-focusing Nonlinear Schr\"odinger (NLS) equation in $1+1$ dimensions
\beq\label{NLS}
i u_t +u_{xx}+2 |u|^2 u=0, \ \ u=u(x,t)\in\CC, 
\eeq
is the simplest universal model in the description of the amplitude modulation of quasi monochromatic waves in weakly nonlinear media; in particular, it is relevant in water waves \cite{Zakharov,AS}, in nonlinear optics \cite{Solli,Bortolozzo,PMContiADelRe}, in Langmuir waves in a plasma \cite{Malomed}, and in the theory of Bose-Einstein condensates \cite{Bludov,Pita}. Its homogeneous solution $u_0(t)=\exp(2i t)$, describing Stokes waves \cite{Stokes} in a water wave context, a state of constant light intensity in nonlinear optics, and a state of constant boson density in a Bose-Einstein condensate, is unstable under the perturbation of waves with wave number $k<2$ \cite{Bespalov,BF,Zakharov} and growth rate $\sigma(k)=k\sqrt{4-k^2}$, and this modulation instability (MI) is considered as the main cause for the formation of anomalous (rogue, extreme, freak) waves (AWs) in nature \cite{KharifPeli3,Onorato2}. The integrability of \eqref{NLS} \cite{ZakharovShabat} allows one to construct a large zoo of exact AW solutions, like the Akhmediev breather (AB) \cite{Akhmed0}
\beq\label{Akhmed}
\ba{l}
e^{2it}{\cal A}(x-x_0,t-t_0,\phi), \ \ \ \   
{\cal A}(x,t,\phi):=\frac{\cosh[\sigma(k) t+2i\phi]+\sin(\phi)\cos(k x)}{\cosh(\sigma(k) t)-\sin(\phi)\cos(k x)},\\
k=2\cos\phi, \ \ \sigma(k)=k\sqrt{4-k^2}=2\sin(2\phi),
\ea
\eeq
where $x_0,t_0$ are arbitrary real parameters associated with the space/time translation symmetries of \eqref{NLS}, space periodic and exponentially localized in time over the background $u_0$, and its long wave limit, the Peregrine instanton \cite{Peregrine}
\beq\label{Peregrine}
e^{2it}{\cal P}(x-x_0,t-t_0), \ \ \ \ {\cal P}(x,t):=1-\frac{4+16it}{1+4 x^2+16 t^2},
\eeq
rationally localized in space-time over $u_0$, together with their multi-mode generalizations \cite{ItsRybinSall}, \cite{Dubard}. Peregrine and Akhmediev AWs are the simplest exact solutions describing the basic features of AWs in $1+1$ dimensions: an exponential growth, due to MI, from the unstable background $u_0(t)$, reaching the amplitude maximum in $(x_0,t_0)$, and the subsequent decay to the background, and they have been observed in many experiments; see, for instance: \cite{CHA_observP,KFFMDGA_observP,Yuen3,Kimmoun,Mussot,Pierangeli}. See \cite{DegaLS,Caso_DegaLS} for the study of the linear instability of background solutions in coupled NLS equations.

In the NLS Cauchy problem of AWs for localized initial perturbations of the background $u_0(t)$, slowly modulated periodic oscillations described by the elliptic solution of \eqref{NLS} play a relevant role in the longtime regime \cite{Biondini1,Biondini2}. If the perturbations are periodic with period $L$, the finite-gap method \cite{Novikov,Dubrovin,ItsMatveev,Lax,MKVM,Krichever} has been adapted in \cite{GS1,GS3} to solve at leading order and in terms of elementary functions the NLS Cauchy problem of the AWs, in the case of a finite number $N=\lfloor L/\pi\rfloor$ of unstable modes. In the simplest case of a single unstable mode $k=2\pi/L$ ($N=1 \ \Leftrightarrow \ \pi<L<2\pi$), one obtains an ideal Fermi-Pasta-Ulam-Tsingou (FPUT) recurrence of ABs, giving a quantitative description in terms of elementary functions of the AW recurrence observed in real and numerical experiments  \cite{Yuen1,Yuen3,Kimmoun,Mussot,Pierangeli}. This one-mode FPUT recurrence can also be described by a simpler matched asymptotic expansion (MAE) approach \cite{GS2}, not requiring the powerful but more technical tools of integrability. See also \cite{GS4} for a finite-gap model describing the numerical instabilities of the AB, and \cite{GS5} for the analytic study of the linear, nonlinear, and orbital instabilities of the AB within the NLS dynamics. See \cite{GS6} for the analytic study of the phase resonances in the AW recurrence, and see \cite{San}, \cite{CS2} and \cite{CS4} for the analytic study of the FPUT AW recurrence in other NLS type models: respectively the PT-symmetric NLS equation \cite{AM1}, the Ablowitz-Ladik lattice \cite{AL}, and  the massive Thirring model \cite{Thirring}. To explain quantitatively the $O(1)$ effects of a small loss or gain on the NLS AW recurrence observed in real and numerical experiments \cite{Kimmoun,Soto}, a finite gap perturbation theory of AWs has been successfully constructed in \cite{CGS1} (see also \cite{CGS2}), and then applied in \cite{CS1} to the Ginsburg-Landau \cite{Newell_Whitehead} and Lugiato-Lefevre \cite{LL} models, viewed as perturbations of NLS, and in \cite{CS3} to hamiltonian and non hamiltonian perturbations of the Ablowitz-Ladik lattice.

In multidimensions, like in the ocean and in the nonlinear optics of crystals, the large majority of physically relevant NLS type models are non integrable, and it is not clear if the NLS AW solutions play any role. Multidimensional (generalizations of) NLS (MNLS) are usually obtained replacing $\partial_x^2$ by $\partial_x^2+a \partial_y^2+b \partial_z^2+\dots, \ a,b\in\RR$. The simplest examples are the following non integrable elliptic NLS (ENLS) and hyperbolic NLS (HNLS) equations in $2+1$ dimensions:
\beq\label{ENLS_HNLS_2+1}
  \ba{l}
      iu_t+u_{xx}+\eta\, u_{yy}+2 |u|^2 u=0, \ \ \eta=\pm 1,
      \ea
    \eeq
respectively for $\eta=1$ and $\eta=-1$, relevant respectively in Kerr optical media \cite{Kelley} and in the study of surface waves in deep water \cite{Zakharov}, their $3+1$ dimensional generalizations
\beq\label{ENLS_HNLS_3+1}
\ba{l}
iu_t+u_{xx}+\eta_1 u_{yy}+\eta_2 u_{zz}+2 |u|^2 u=0, \ \ \eta_j=\pm 1, \ \ j=1,2,
\ea
\eeq 
elliptic for $\eta_j=1$, and hyperbolic in the two cases $\eta_j=(-1)^j$ and $\eta_j=-1$, $j=1,2$, together with families of $2+1$ dimensional Davey-Stewartson (DS) type equations \cite{Benney,DS,ABB}, in which the complex amplitude $u$ is coupled with the real mean flow potential.

The guiding principle of our research on multidimensional AWs is based on the common sense argument of the lighthouse: the navigation in a dark night near the coast is difficult/dangerous, and the presence, here and there, of lighthouses is a great help .... When studying MNLS equations we have two lighthouses; the first one is associated with the existence of integrable DS equations \cite{DS,Shulman}, and in \cite{GS7} a finite gap theory for doubly periodic AWs of the integrable DS2 equation has been constructed, allowing one to solve the periodic Cauchy problem of AWs at leading order in terms of elementary functions, like in the NLS case; while in \cite{CGS3} the $N$-breather AW solution of Akhmediev type of the integrable DS1 and DS2 equations has been constructed, playing a key role in the description of the AW recurrence, and in the phenomenological study of AWs of integrable and non integrable MNLS equations \cite{CS5}.

The second lighthouse, common to integrable and non integrable MNLS equations, is the quasi one dimensional (Q1D) regime in which the wavelength in the direction of propagation $x$ is small with respect to the wavelengths in the transversal directions
\beq
\frac{\lambda_x}{\lambda_y},\frac{\lambda_x}{\lambda_z},\dots =O(\delta), \ \ \delta\ll 1 ,
\eeq
or, equivalently, the dependence on the extra space variables $y,z,\dots$ is slow. In this regime all the above MNLS equations are close to NLS, and one expects that suitable adiabatic deformations of the AB \eqref{Akhmed} and of the Peregrine instanton \eqref{Peregrine} be relevant. The goal of this paper is to uncover the universal analytic features of AWs of MNLS equations in the Q1D regime, concentrating our attentions on the first nonlinear stage of MI (NLSMI), and showing, at leading order, a universal behavior, independent of the particular model, and described in terms of suitable adiabatic deformations of the AB \eqref{Akhmed}. Varying the initial data, the first NLSMI consists in various combinations of the following two basic processes. P1: the AW growth from the unstable background, followed by fission in the transversal slowly varying directions, with infinite speed at fission time. P2: the inverse process of fusion in the transversal slowly varying directions, with infinite speed at fusion time, followed by AW decay to the background. Fission and fusion are critical processes showing similarities to multidimensional wave breaking, and to phase transitions of second kind and critical exponent $1/2$. They may or may not coexist in the same nonlinear stage of MI, but while the universal process of ``AW growth + fission'' always occurs, fusion is less generic. We also show that, in the natural long wave limit, the above Q1D Akhmediev breather reduces to the Q1D analogue of the Peregrine AW, rationally localized in the space variables, also undergoing the above critical processes. At last we show, through numerical experiments on the HNLS equation, that the process of ``AW growth + fission'' is by no means restricted to the Q1D regime, extending to a finite region of the MI domain of the equation (see \cite{CS5} for more details). The recurrence of Q1D AWs, another important feature of $x$-periodicity, turns out to be different for different MNLS models, and the quantitative description of these differences is postponed to the subsequent paper \cite{CS7}. Due to the universality of the above processes, it is very plausible that they be observable in natural phenomena in which AWs of MNLS equations be relevant, like in water waves, nonlinear optics, plasma physics, and Bose-Einstein condensates. 

\section{AW fission and fusion}
 
We first concentrate on Q1D AWs in $2+1$ dimensions, choosing as reference models equations \eqref{ENLS_HNLS_2+1}, and on initial perturbations of the background  periodic in $x$, with period $L_x$ corresponding to a single unstable mode: $\pi<L_x<2\pi$, and slowly varying and even functions of $y$:
\beq\label{Cauchy_data}
  \ba{l}
u(x,Y,0)=1+\eps \Big[c_{+}(Y)e^{i k_x x}+c_{-}(Y)e^{-i k_x x}+\mbox{stable part of the Fourier} \\
x\mbox{-series}\Big], \ \ k_x=\frac{2\pi}{L_x}, \ \ c_{\pm}(-Y)=c_{\pm}(Y), \ \ Y=\delta y, \ \ \eps,\delta\ll 1,
\ea
\eeq
and we follow the MAE approach in \cite{GS2}. For the application of this approach to the study of the NLS evolution of localized slowly varying envelops with carrier wave number $k$: $u(x,0)=1+\eps [c_+\!(\delta x)\exp(ikx)+c_-\!(\delta x)\exp(-ikx)]$, see \cite{Trillo2}.

The linear stage of MI, for $0\le t \le O(1)$, described by $u=e^{2it}(1+\eps w)$ and $iw_t-2w+w_{xx}\pm\delta^2w_{YY}+4w+2 w^*=0$, where $w^*$ is the complex conjugate of $w$, is the same for both equations \eqref{ENLS_HNLS_2+1} and, in general, for all MNLS equations, up to $O(\delta^2)$ corrections, and coincides with the well known one of NLS \cite{GS1,GS2} treating the slow variable $Y$ as parameter:
\beq\label{linear_stage}
\ba{l}
u_{lin}(x,Y,t)= e^{2it}\Big[1+\eps\Big(\frac{2|\alpha(Y) |}{\sigma(k_x)}e^{\sigma(k_x) t+i\phi}\cos[k_x (x-x_1(Y))]\\
+ \frac{2|\beta(Y) |}{\sigma(k_x)}e^{-\sigma(k_x) t-i\phi}\cos[k_x (x-x_0(Y))]\Big)+O(\eps)\mbox{ oscillations} \Big)\Big],
\ea
\eeq
where $\phi=\arccos(k_x/2)$, $\sigma(k_x)=k_x\sqrt{4-k_x^2}=2\sin(2\phi)$, and
\begin{equation}\label{alpha_beta}
  \ba{l}
x_1(Y)=\frac{\arg(\alpha(Y))+\pi/2}{k_x}, \ \ x_0(Y)=\frac{-\arg(\beta(Y))+\pi/2}{k_x}, \\  
\alpha(Y) =e^{-i\phi_1}c^*_+(Y)-e^{i\phi_1}c_{-}(Y), \ \  \beta(Y) =e^{i\phi_1}c^*_{-}(Y)-e^{-i\phi_1}c_+(Y).
\ea
\end{equation}
Then, for $1\ll t \ll \frac{1}{\sigma}\ln\left(\frac{1}{\eps}\right)$, we have the asymptotics
\beq\label{overlapping1}
u(x,Y,t)\sim e^{2it}\Big[1+\frac{2\epsilon|\alpha(Y) |}{\sigma(k_x)}e^{\sigma(k_x) t+i\phi}\cos(k_x (x-x_1(Y)))\Big],
\eeq
matching successfully with the following adiabatic deformation of the AB \eqref{Akhmed} \cite{GS2}
\beq\label{1st_appearance}
\ba{l}
u_1(x,Y,t)={\cal A}\left(x-x_1(Y),t-t_1(Y),\phi\right)e^{2it+2i\phi},\\
t_1(Y)=\frac{1}{\sigma(k_x)}\log\left(\frac{\sigma^2}{2 \eps |\alpha(Y)|}\right), \ \ |t-t_1(Y)| =O(1).
\ea
\eeq
Equation \eqref{1st_appearance} describes at leading order the first NLSMI, the same for equations \eqref{ENLS_HNLS_2+1} and for all  MNLS equations, up to small corrections whose analytic investigation is postponed to a subsequent paper; therefore it describes the universal features of Q1D AWs. Also for the ENLS equation, well known to lead to a blow up of generic initial perturbations of $u_0$ \cite{Sulem}, the first NLSMI is described by \eqref{1st_appearance} if the MI time scale $\ln(1/\eps)$ is much smaller than the Q1D scale $1/\delta$, implying that blow up occurs only after a number of recurrences.  

\begin{figure}[h!!!!]
	\centering
	\includegraphics[trim=0cm 1cm 0cm 0 ,width=13cm,height=4cm]{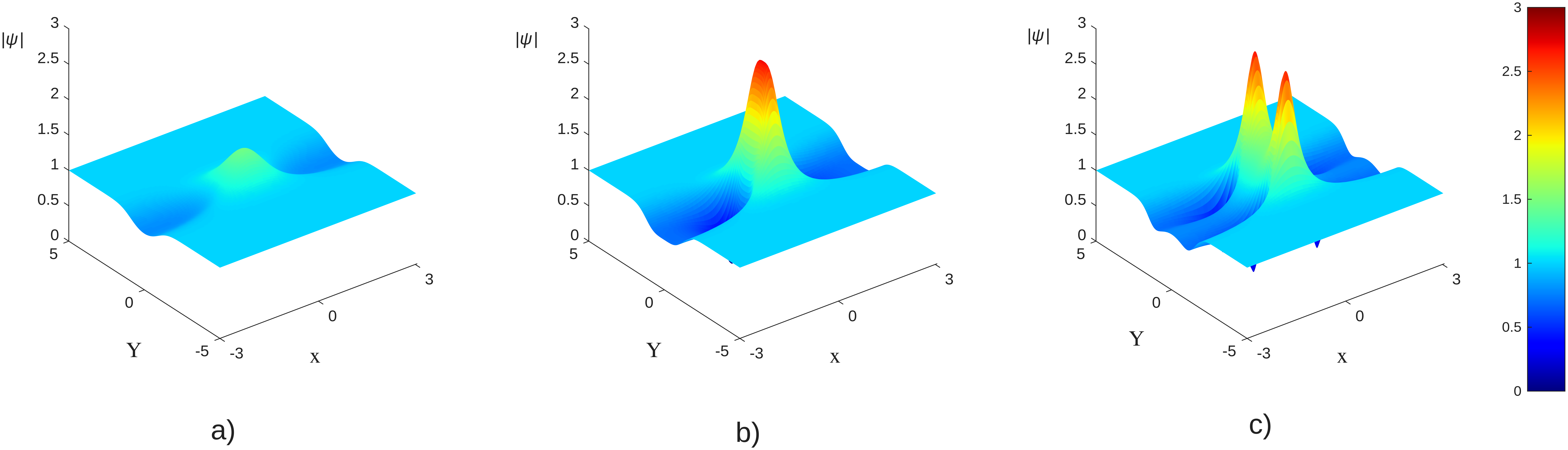}
	\caption{Numerical evolution according to the HNLS equation for the initial condition $c_{\pm}(Y)=e^{-Y^2}/2$, $L_x=6$, $\eps=\delta=10^{-2}$, focusing on the first appearance and subsequent fission. \textbf{a)} t=2.97: AW growth.  \textbf{b)} The fission time $t_{fiss}=3.58$.  \textbf{c)} $t=4.15$ the separation of the fission products.  }\label{fission}
\end{figure}

To understand the main properties of the first NLSMI \eqref{1st_appearance}, we restrict the family of initial data in \eqref{Cauchy_data} to
\beq\label{FourierCoeff_2+1}
c_{\pm}(Y)=c_{0\pm}f(Y)e^{\mp i g(Y)},
\eeq
where $c_{0\pm}$ are arbitrary complex constants, and $f(Y),g(Y)$ are real even functions of $Y$, assuming without loss of generality that $0< f(Y)\le 1$, implying
\beq\label{parameters_2+1}
\ba{l}
\alpha(Y)=\alpha_0 f(Y)e^{i g(Y)}, \ \ \beta(Y)=\beta_0 f(Y)e^{-i g(Y)}, \\
\alpha_0=e^{-i\phi}c^*_{0+}-e^{i\phi}c_{0-}, \ \ \beta_0=e^{i\phi}c^*_{0-}-e^{-i\phi}c_{0+}, \\
t_1(Y)=t_0+\frac{1}{\sigma(k_x)}\log\left(\frac{1}{f(Y)} \right), \ \ t_0=\frac{1}{\sigma(k_x)}\log\left(\frac{\sigma^2}{2\eps |\alpha_0 |} \right), \\
x_1(Y)=x_{10}+\frac{g(Y)}{k_x}, \ \ x_{10}=\frac{\arg\alpha_0+\pi/2}{k_x},
\ea
\eeq
and we begin with the simplest case  of a single hump (in $Y=0$) function $f$: 
\beq\label{about_Y=0_2+1}
\ba{l}
f(Y)=1-\frac{|f''(0)|}{2}Y^2 +O(Y^4), \ |Y|\ll 1, \\
g(Y)=d\, Y^2+O(Y^4), \ |Y|\ll 1, \ d\in\RR. \\
\ea\eeq
Then $t_1(Y)$ has a single minimum $t_0$ at $Y=0$:
\beq
\ba{l}
t_1(Y)=t_0+\frac{|f''(0)|}{2\sigma(k_x)}Y^2+O(Y^4), \ \ |Y|\ll 1, \\
x_1(Y)=x_{10}+\frac{d}{k_x} Y^2+O(Y^4), \ |Y|\ll 1,
\ea
\eeq
and since \eqref{1st_appearance} reaches its amplitude maximum along the curves $x=x_1(Y)$ and $t=t_1(Y)$, the first maximum is when $t_1(Y)$ is minimum, namely at $(x,Y,t)=(x_{10},0,t_0)$, with $|u_1|=1+2\sin\phi$. At $t_0$ the AW \eqref{1st_appearance} splits into two waves of amplitude $1+2\sin\phi$, traveling on the trajectories $x=x_1(Y_{\pm}(t))$, where $Y_{\pm}(t)=\pm\sqrt{\frac{2\sigma(k_x)}{|f''(0)|}(t-t_0)}, \ 0\le t-t_{0}\ll 1$ (see Figure \ref{fission}). Therefore the fission time is $t_{fiss}=t_0$, and the two fission products separate with infinite speed at fission time according to the universal law:
\beq\label{fission_speed_2+1}
\dot{Y}_{\pm}=\pm \sqrt{\frac{\sigma(k_x)}{2|f''(0)| }}\frac{1}{\sqrt{t-t_{fiss}}}, \ \ 0\le t-t_{fiss}\ll 1.
\eeq

\begin{figure}[h!!!!]
	\centering
	\includegraphics[trim=0cm 1cm 0cm 0 ,width=13cm,height=4cm]{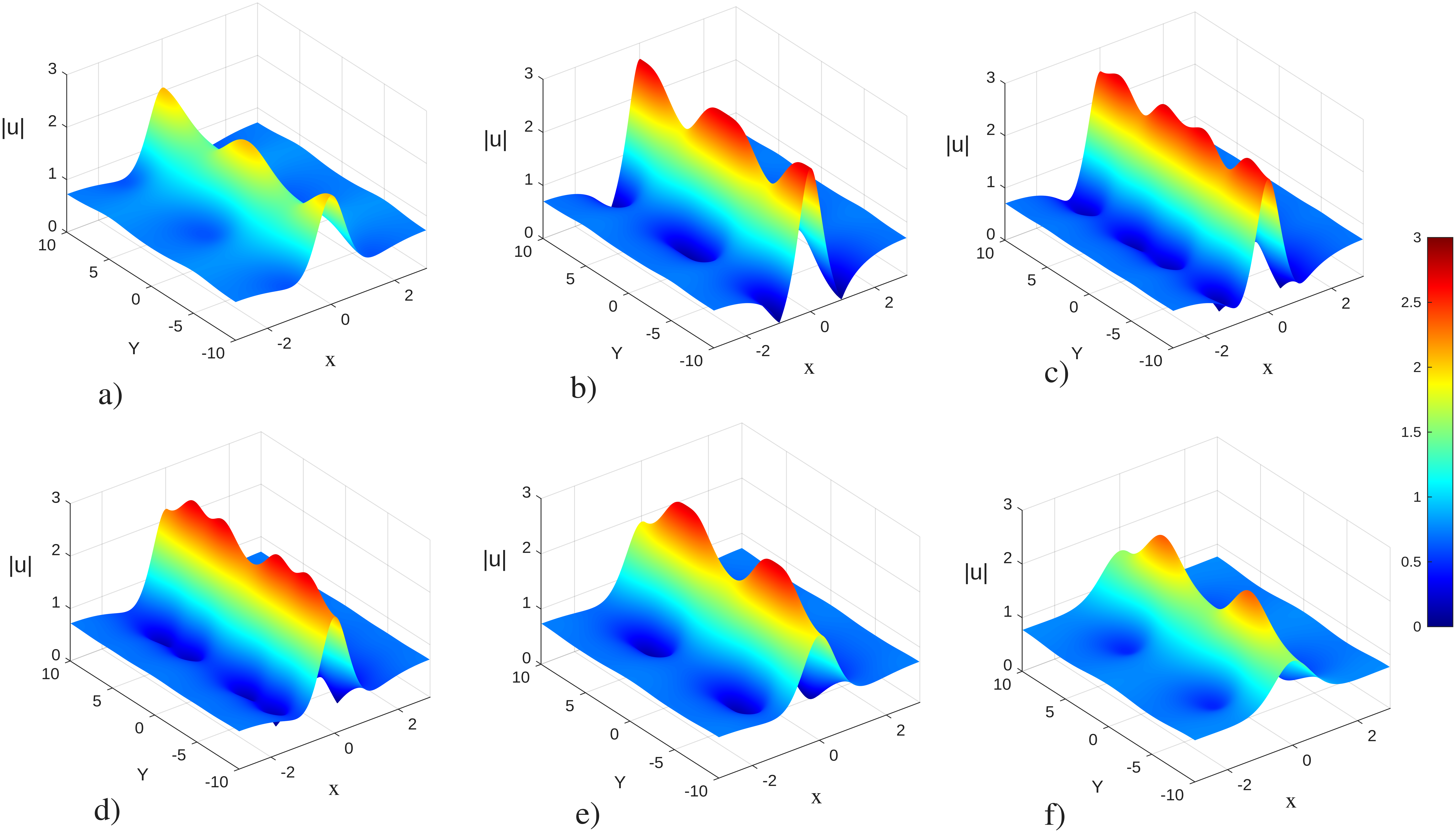}
	\caption{Six snapshots describing the first NLSMI corresponding to the initial condition \eqref{FourierCoeff_2+1} with $c_{0-}=0.3+i 0.1$, $c_{0+}=0.084$, $f(Y)=\frac{1+0.3\cos(0.628\, Y)}{1.3}$, $g(Y)=0$, $\eps=10^{-2}, \ \delta=10^{-3}$, and $x\in [-L_x/2,L_x/2]$, $Y\in [-L_Y,L_Y]$, with $L_x=6, \ L_Y=10$. a) $t=3.3$: the growth from the background. b) $t=t_{fiss}=3.59$: the fission in $(x_{10},Y_n)$, $n=0,\pm 1$; c) $t=3.70$: the separation of the fission products. d) $t=3.82$ the right fission product in $0$ and the left fission product in $Y=L_y$ travel against each other and get closer; e) $t=t_{fus}=3.95$: the fusion in $(x_{10},\pm L_Y/2)$. f) $t=4.13$: the AW decay to the background.}\label{1st_fission_cos}
\end{figure}

\begin{figure}[h!!!!]
	\centering 
	\includegraphics[trim=0cm 1cm 0cm 0 ,width=6.5cm,height=4cm]{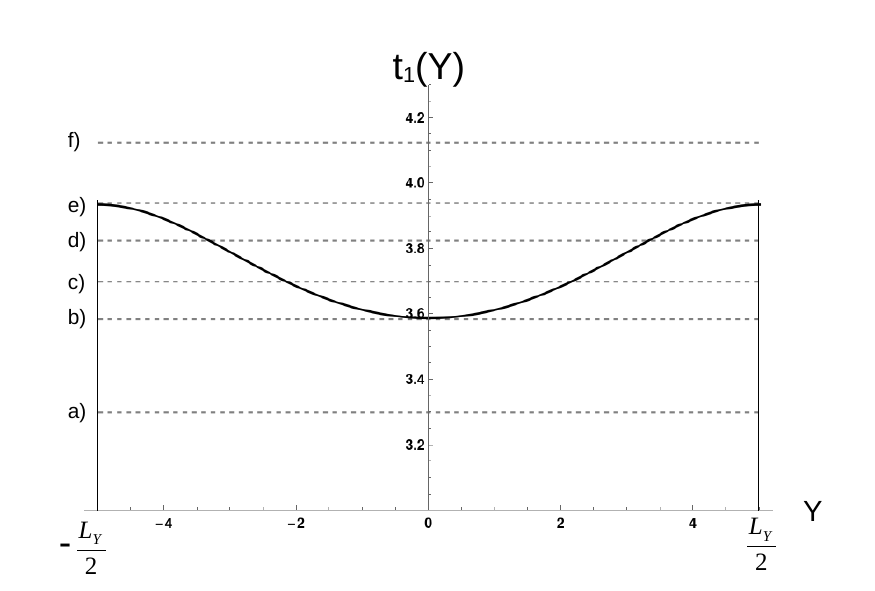}
        \includegraphics[trim=0cm 1cm 0cm 0 ,width=6.0cm,height=4cm]{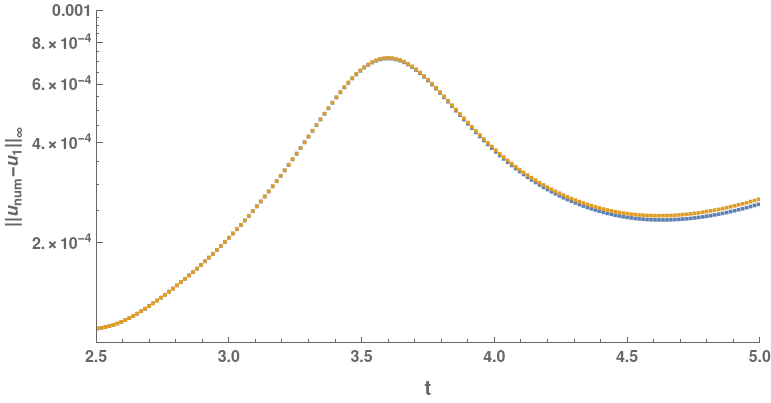}
	\caption{Left: the plot of $t_1(Y)$ as function of $Y$; the horizontal dashed lines correspond to the six snapshots of Figure \ref{1st_fission_cos}. Right: the uniform distance \eqref{uniform_distance} in orange and in blue respectively for the ENLS and the HNLS equation.}\label{moto_ty_cos}
\end{figure}

As time grows, the behavior of the two fission products depends on the behavior of $f(Y)$ far from $Y=0$. Here we consider the following cases.
\vskip 3pt
\noindent
1) If $f(Y)$ is even and periodic of period $L_Y$, with a single hump inside the period, like if $f(Y)=\frac{1+\gamma\cos(2\pi Y/L_Y)}{1+\gamma}, \ 0<\gamma<1$,  and $g(Y)=0$, then $t_1(Y)$ has minima in $Y_n=n L_Y$, with $t_1(Y_n)=t_0$, and maxima in $\tilde Y_n=(n+1/2)L_Y$, with $t_1(\tilde Y_n)=t_0+\ln\!\left(\frac{1+\gamma}{1-\gamma} \right)$, $n\in\ZZ$ (see the left Figure \ref{moto_ty_cos}). Consequently growing AWs undergo fission at $t_{fiss}=t_0$ in $(x_{10},Y_n)$; then the right fission product in $Y_n$ and the left fission product in $Y_{n+1}$ travel against each other along the straight line $x=x_{10}$, and undergo fusion in $(x_{10},\tilde Y_n)$ at $t_{fus}=t_1(\tilde Y_n)$ into an AW decaying to the background (see Figure \ref{1st_fission_cos} and the left Figure \ref{moto_ty_cos}). To evaluate how well \eqref{1st_appearance} describes this NLSMI, we plot in the right Figure \ref{moto_ty_cos} the uniform distance between the numerical evolution and the analytic approximant \eqref{1st_appearance} as function of time:
\beq\label{uniform_distance}
\| u_{num}-u_1\|_{\infty}(t):=\!\!\!\sup_{x\in [0,L_x], y\in [0,L_Y]}\!\!\!\!\!|u_{num}(x,Y,t)-u_1(x,Y,t) |,
\eeq
for both ENLS and HNLS equations, in a time interval containing both fission and fusion, obtaining an excellent agreement: the two distances are essentially indistiguishable, and their max is less than $8\cdot 10^{-4}$, for $\eps=10^{-2}, \delta=10^{-3}$.    
\vskip 3pt
\noindent
2) If $f(Y)=1/\cosh(Y)$ and $g(Y)=d\, Y^2$, then at $t_{fiss}=t_0$ the AW splits into two waves traveling along the parabola $k(x-x_{10})=d\, Y^2_{\pm}(t)$, where $Y_{\pm}(t)=\pm \ln\Big(e^{\sigma(t-t_0)}+\sqrt{e^{2\sigma(t-t_0)}-1} \Big), \ t-t_0>0$ in opposite directions (if $d=0$, they travel along the straight line $x=x_{10}$ parallel to the $Y$ axis). When $t-t_{fiss}\gg 1$, the two fission products are well separated and described at leading order by
\beq\label{u+-_sech}
  \ba{l}
  u^{\pm}_1= \frac{\cosh\left[Y\mp \left(\sigma(t-t_0)+\ln 2\right)\mp 2i\phi\right]+\sin\phi \cos\left[k_x(x-x_{10})-d\, Y^2\right]}
  {\cosh\left[Y\mp \left(\sigma(t-t_0)+\ln 2\right)\right]-\sin\phi \cos\left[k_x(x-x_{10})-d\, Y^2\right]}e^{2it+2i\phi},  \ t-t_{fiss}\gg 1.
  \ea
  \eeq
It follows that they travel asymptotically with constant speeds $\pm \sigma(k_x)$ without changing their shape, and no fusion takes place.
\vskip 3pt
\noindent
3) If the localization is faster, like for $f(Y)=e^{-Y^2}$, and $g(Y)$ is again $d\, Y^2$, then at $t_{fiss}=t_0$ the AW splits into two waves traveling along the parabola $k(x-x_{10})=d\, Y^2_{\pm}(t)$, but now $Y_{\pm}(t)=\pm\sqrt{\sigma(t-t_0)}$ in opposite directions (if $d=0$, they travel along the straight line $x=x_{10}$). As $t-t_{fiss}\gg 1$, the two fission products are well separated and described now by
  \beq\label{u+-_gauss}
  \ba{l}
  u^{\pm}_1 = \frac{\cosh\left(2\sqrt{\sigma(t-t_0)}\xi^{\pm}\mp 2i\phi\right)+\sin\phi \cos(k(x-x_{10}-d\, Y^2))}
  {\cosh\left(2\sqrt{\sigma(t-t_0)}\xi^{\pm}\right)-\sin\phi \cos(k(x-x_{10}-d\, Y^2))}e^{2it+2i\phi}, \\
  \xi^{\pm}=Y\mp\sqrt{\sigma(t-t_0)}, \ \ t-t_0\gg 1.
  \ea
  \eeq
It follows that their speed and width in the $Y$ direction decrease like $(t-t_{fiss})^{-1/2}$ as $t-t_{fiss}$ increases, and, due to this shrinking, at some point the Q1D approximation, as well as the multiscale expansion hypothesis under which the models \eqref{ENLS_HNLS_2+1} are derived, cease to be valid.
\vskip 3pt
\noindent
4) If $f(Y)$ possesses more local maxima and minima, the NLSMI becomes richer, combining fission and fusion events, and here we illustrate just the case of an even and localized function with two maxima at $\pm Y_M$ and a minimum in $0$, with $g(Y)=0$. Then $t_1(Y)$ has two equal minima in $\pm Y_M$ and a maximum in $0$, and the first NLSMI consists of the following steps. The simultaneous growth of two AWs undergoing fission in $(x_{10},\pm Y_M)$ at $t_{fiss}=t_1(Y_M)$. While the external fission products separate in opposite directions, without undergoing fusion, the two internal fission products travel towards each other and undergo fusion in $(x_{10},0)$ at $t_{fus}=t_0$, with a subsequent decay to the background. It is possible to verify also for the examples 2)-4) that the uniform distance \eqref{uniform_distance} is very small.

The above examples of NLSMI show an intrinsic instability of the growing Q1D AW, leading to its fission. On the other hand, the fission products, whose nature is different from that of the growing AW (compare \eqref{1st_appearance} with \eqref{u+-_sech},\eqref{u+-_gauss}), have an intrinsic tendency to undergo fusion into an AW decaying to the background, but only if the geometry of the process favours their frontal collision, as in the examples 1) and 4) above. In addition fusion requires the previous creation of fission products, and only special initial data lead to AW fusion without prior AW fission. 

\begin{figure}[h!!!!!!!!!!!!!!!]
	\centering
	\includegraphics[width=9.0cm]{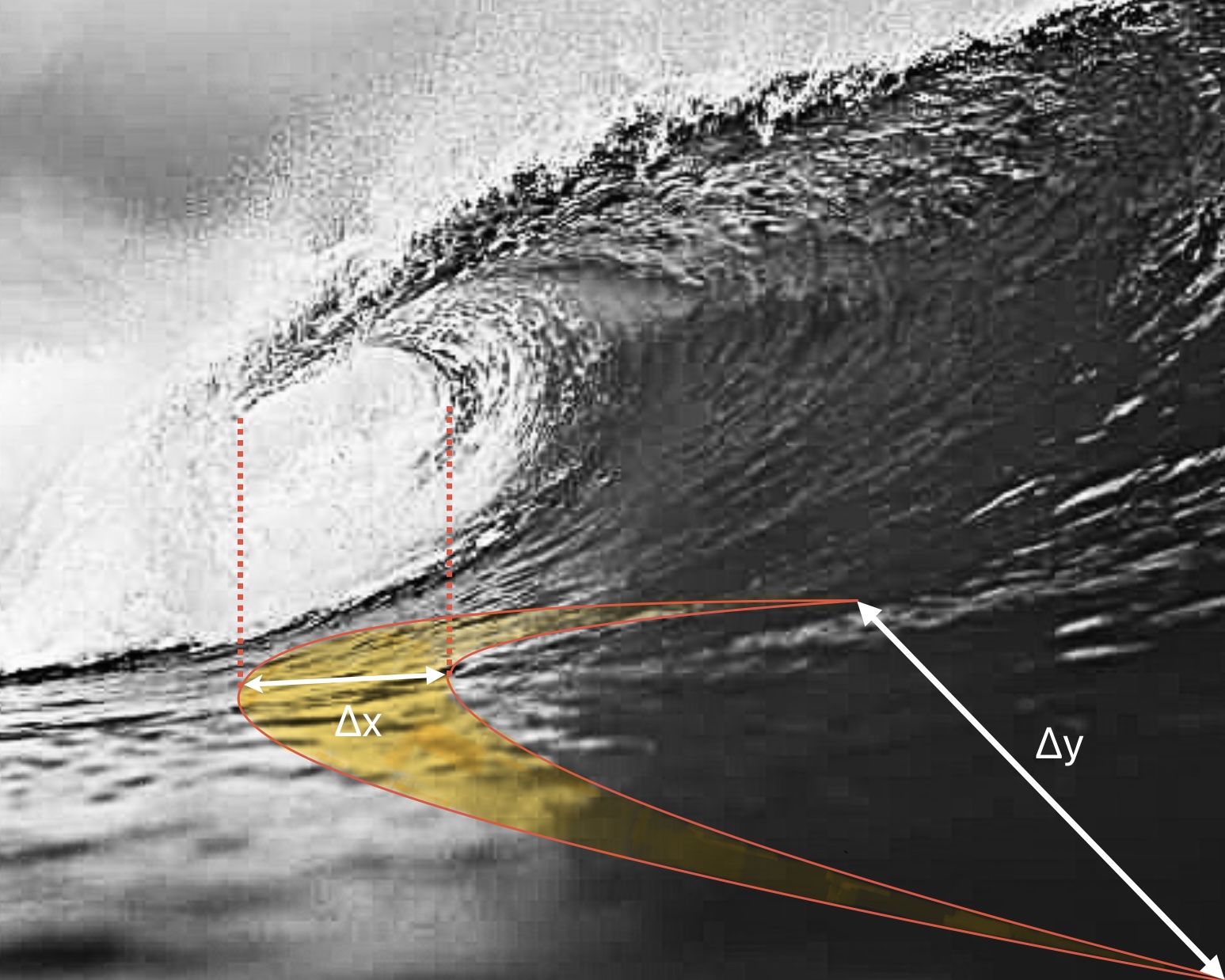}
\caption{A $2+1$ dimensional wave breaking, and the domain (in yellow) in the $(x,y)$ plane in which the elevation is a three-valued function.} \label{breaking}	
\end{figure}

\begin{figure}[h!!!!!!!!!!!!!!!]
	\centering
	\includegraphics[width=11.0cm]{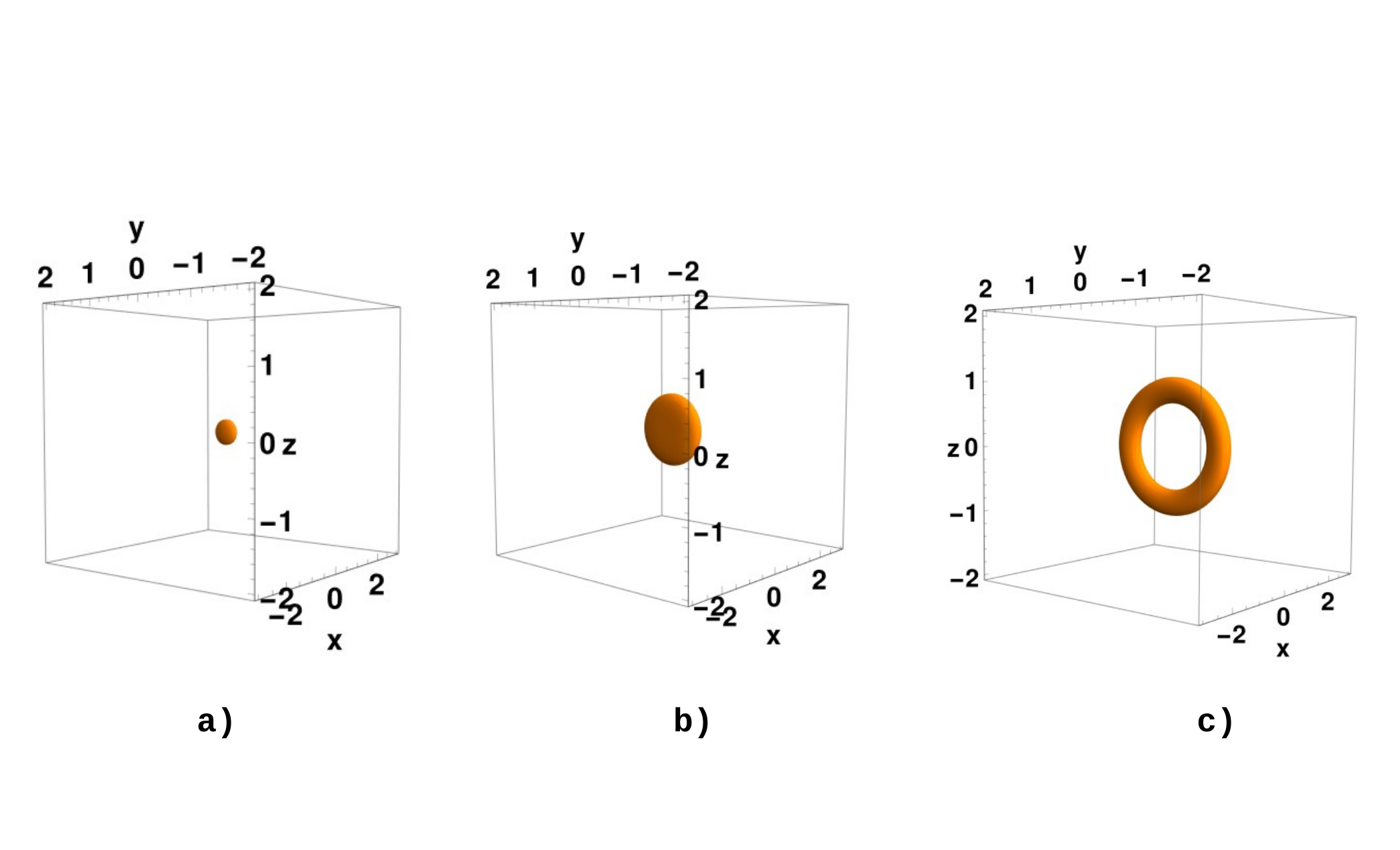}
\caption{The numerical evolution of $|u_1(x,R,t)|$ in the first NLSMI for equation \eqref{ENLS_HNLS_3+1} with $\eta_1=\eta_2=-1$, and $c_{\pm}=e^{-R^2}/2$, $\eps=\delta=10^{-2}$, $L_x=6$, plotting only regions in which $|u_1|>2.2$, for the sake of clarity.  a) $t=2.75$: AW growth from the background. b) $t=t_{fiss}=2.93$: the fission time. c) $t=3.4$: the opening smoke ring centered on the $x$ axis.} \label{3D_smoke_ring}	
\end{figure}

\begin{figure}[h!!!!!!!!!!!!!!!]
	\centering
	\includegraphics[width=10.0cm]{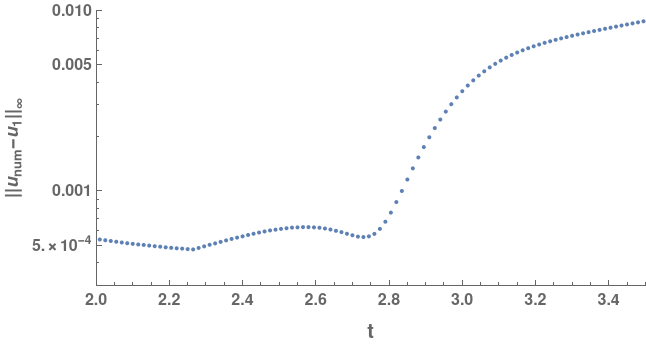}
\caption{The uniform distance between the numerics of the experiment in Figure \ref{3D_smoke_ring} and the theoretical approximant \eqref{1st_appearance_3+1}, in a time interval containing the snapshots of Figure \ref{3D_smoke_ring}. The distance is less than $8\cdot 10^{-3}$.} \label{3D_diff}	
\end{figure}

\begin{figure}[h!!!!!!!!!!!!!!!]
	\centering
	\includegraphics[width=13.0cm]{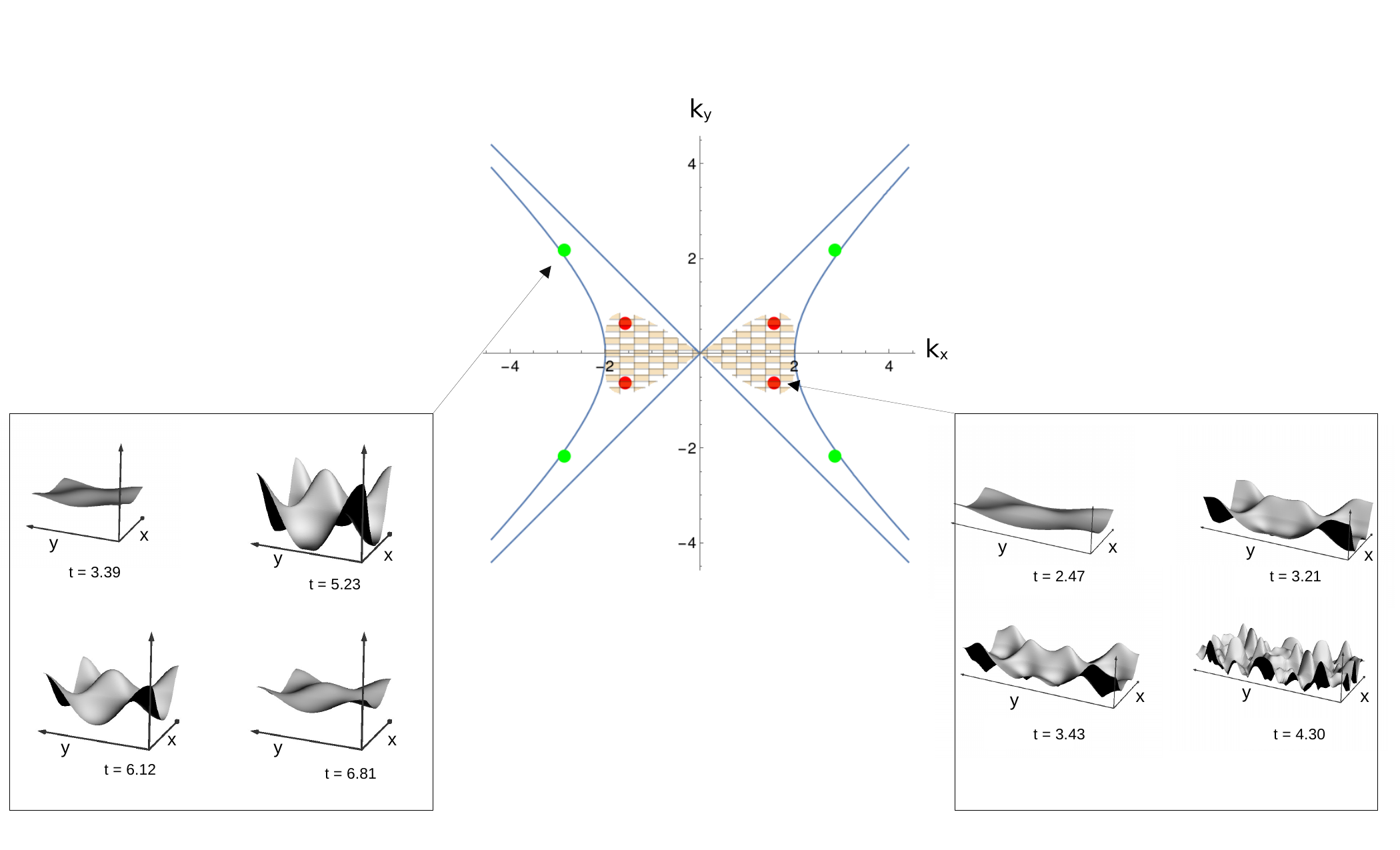}
\caption{The central figure shows the MI domain $0<k^2_x-k^2_y<4$ of the HNLS equation. The right figure shows four snapshots of the HNLS numerical evolution of a doubly periodic initial condition $1+\eps(e^{i\phi}\cos(k_x x)\cos(k_y y))$, with periods $L_x=4, \ L_y=10$, where $k_x=2\pi/L_x, \ k_y=2\pi/L_y$ are the coordinates of the red points of the central figure, and $\phi=\arccos(\sqrt{k^2_x-k^2_y}/2)$; this AW undergoes fission. The left figure shows four snapshots of the HNLS numerical evolution for the same initial condition, but with different periods $L_x=2.2, \ L_y=2.89$, corresponding to the green points of the central figure; this AW does not undergo fission. } \label{HNLS_fission_region}  	
\end{figure}

Now we show how the basic NLSMI in $1+1$ dimensions: ``growth from the background and subsequent decay to the background'', well expressed by the formula $\sign(\rho_t(t))=-\sign(t-t_0)$, $|t-t_0|\ll 1$, where $\rho(t)=|{\cal A}(0,t-t_0)|^2$, implies the process of ``growth from the background + fission'' (and its inverse) for Q1D AWs. We begin with formulas
\beq
\ba{l}
{u_1}_{Y}=-\left(\dot x_1 {\cal A}_x+\dot t_1 {\cal A}_t\right)e^{2it+2i\phi}, \\
{u_1}_{YY}=\left(-\ddot x_1{\cal A}_x -\ddot t_1{\cal A}_t +2 \dot x_1\dot t_1{\cal A}_{xt}+{\dot x_1}^2{\cal A}_{xx}+{\dot t_1}^2{\cal A}_{tt}\right)e^{2it+2i\phi}.
\ea
\eeq
Since $\dot x_1(0)=\dot t_1(0)=0$, and $\ddot t_1(0)>0$, it follows that, at the max point $(x,Y)=(x_{10},0)$, ${u_1}_{Y}={u_1}_{x}=0$, and ${u_1}_{YY}(x_1(0),0,t)=-\ddot t_1(0){\cal A}_t(0,t-t_0)$, $|t-t_0|\ll 1$. Consequently $(|u_1(x_{10},0,t) |^2)_{YY}=-\ddot t_1(0)\rho_t(t)$, implying that $\sign\!\left((|u_1(x_{10},0,t) |^2)_{YY}\right)=\sign (t-t_0)$, $|t-t_0|\ll 1$. It follows that, when the Q1D AW reaches it max, its $Y$-curvature changes sign, from negative to positive sign, corresponding to fission in the $Y$ direction.
\vskip3pt
\noindent
\textbf{Fission and fusion as critical processes}. Fission and fusion are critical processes showing similarities i) to multidimensional wave breaking according to the integrable dispersionless Kadomtsev-Petviashvili (dKP) equation $(u_t+u u_x)_x+u_{yy}=0$ (describing for instance the evolution of small amplitude, nearly one-dimensional waves in shallow water near the shore), and ii) to phase transitions of second kind and critical exponent $1/2$. Indeed, i) in the dKP wave breaking, a smooth localized initial datum undergoes the first wave breaking in a space point $(x_b,y_b)$ at time $t_b$, constructed from the initial data via the spectral transform developed for integrable dispersionless PDEs \cite{MS0,MS1}. For $0<t-t_b\ll 1$ the multivaluedness region develops from the breaking point $(x_b ,y_b)$ with the law $\Delta x=O\left((t-t_b)^{3/2}\right)$ and velocity $\Delta \dot x=O\left((t-t_b)^{1/2}\right)$ (zero speed at breaking time) in the $x$ direction, like in the $1+1$ dimensional reduction $u_t+u u_x=0$ (the Hopf equation), while it develops with the law $\Delta y=O\left((t-t_b)^{1/2}\right)$ and velocity $\Delta \dot y=O\left((t-t_b)^{-1/2}\right)$ (infinite speed at breaking time) in the slowly varying direction \cite{MS1,MS2} (see Figure \ref{breaking}). ii) In the Landau-Ginzburg theory of phase transitions, the behavior of the order parameter around the critical temperature $T_c$ is a two-valued function proportional to $\pm\theta(T_c-T)\sqrt{T_c-T}$, where $\theta(\cdot)$ is the step function, and the susceptivity is singular at $T_c$, with critical exponent $1/2$ \cite{Parisi}.
\vskip3pt
\noindent
\textbf{$3+1$ dimensions}. The above considerations generalize in a straightforward way to $3+1$ dimensions for equations \eqref{ENLS_HNLS_3+1}, and for simplicity we assume radial symmetry in the slowly varying transversal plane. Then all above formulas are valid, replacing $Y$ by $R=\sqrt{Y^2+Z^2}, \ Z=\delta z$, and the initial condition  
\beq\label{Cauchy_data_3+1}
  \ba{l}
u(x,R,0)=1+\eps \Big[c_{+}(R)e^{i k_x x}+c_{-}(R)e^{-i k_x x}+\mbox{stable part of the Fourier} \\
\mbox{-series}\Big], \ \ k_x=\frac{2\pi}{L_x}, \ \ R=\sqrt{Y^2+Z^2}, \ \ Y=\delta y, \ \ Z=\delta z, \ \ \eps,\delta\ll 1
\ea
\eeq
leads to the first NLSMI described, at leading order, by
\beq\label{1st_appearance_3+1}
\ba{l}
u_1(x,R,t)={\cal A}\left(x-x_1(R),t-t_1(R),\phi\right)e^{2it+2i\phi},\\
x_1(R)=\frac{\arg(\alpha(R))+\pi/2}{k_x}, \ \ t_1(R)=\frac{1}{\sigma(k_x)}\log\left(\frac{\sigma^2\!(k_x)}{2 \eps |\alpha(R)|}\right),\\
\alpha(R) =e^{-i\phi_1}c^*_+(R)-e^{i\phi_1}c_{-}(R),
\ea
\eeq
with
\beq
\ba{l}
t_1(R)=t_0+\frac{|f''(0)|}{2\sigma}R^2+O(R^4), \ \ R\ll 1, \\
x_1(R)=x_{10}+d\, R^2+O(R^4), \ R \ll 1.
\ea
\eeq
Since the solution \eqref{1st_appearance_3+1} reaches its maximum at $x=x_1(R)$ and $t=t_1(R)$, then the first maximum is at $(x,R,t)=(x_{10},0,t_0)$. At $t_{fiss}=t_0$ the AW \eqref{1st_appearance_3+1} generates a transversal smoke ring, centered on the $x$ axis, opening with growing radius $R(t)=\sqrt{\frac{2\sigma(k_x)}{|f''(0)|}(t-t_0)}, \ 0\le t-t_{fiss}\ll 1$, and living on the rotation surface $x-x_1(R(t))=0$ (on the transversal $(Y,Z)$ plane if $x_1(R)=x_{10}$), see Figure \ref{3D_smoke_ring}. The opening speed is again infinite at $t_{fiss}$, according to the universal law:
\beq\label{fission_speed_2+1}
\dot{R}(t)=\pm \sqrt{\frac{\sigma(k_x)}{2|f''(0)| }}\frac{1}{\sqrt{t-t_{fiss}}}, \ \ 0\le t-t_{fiss}\ll 1.
\eeq
Therefore Q1D AWs in $3+1$ dimensions, radially symmetric in the transversal plane, grow from the background undergoing fission when they reach their maximal amplitude, and fission consists in the generation of an opening smoke ring in the transversal plane. Again the uniform distance \eqref{uniform_distance} between the numerics and the analytic approximant \eqref{1st_appearance_3+1} is very small (see Figure \ref{3D_diff}).
\vskip3pt
\noindent
\textbf{The long wave limit}. The long wave limit of equations \eqref{1st_appearance},\eqref{1st_appearance_3+1} leads to the following adiabatic deformations of the Peregrine solution in $2+1$ dimensions:
\beq\label{Peregrine_2+1}
e^{2it}{\cal P}(x-x_0-d\, y_1^2,t-t_0-a y_1^2), \ \ y_1=\delta^{3/2}y, \ \ a>0, \ \ d\in\RR,
\eeq
and in $3+1$ dimensions with radial symmetry in the transversal plane:
\beq\label{Peregrine_3+1}
e^{2it}{\cal P}(x-x_0-d\, r_1^2,t-t_0-a\, r_1^2), \ \ r_1=\delta^{3/2}\sqrt{y^2+z^2}, \ \ a>0, \ \ d\in\RR,
\eeq
also describing processes P1 and P2. To show it, we first rewrite $Y\to \delta\, k_y\, y$, and choose $k_x=k_y=\delta^p, \ p>0$, implying that $\sigma(k_x)\sim 2\delta^p$. Then $\sigma(k_x)(t-t_1(Y))\sim 2\delta^p(t-t_0-a y_1^2)$ and $k_x(x-x_1(Y))\sim \delta^p(x-x_0-d y_1^2)$, where $y_1=\delta^{(2+p)/2}y$, $d\in\RR$, and $a>0$. Since \eqref{Peregrine_2+1},\eqref{Peregrine_3+1} are the leading order terms of an expansion in integer powers of $\delta$, and since $u_{yy}=\delta^{2+p}u_{y_1 y_1}$ in the ENLS and HNLS equations, we choose $p$ to be the smallest positive number such that $2+p\in\NN^+$, namely $p=1$, obtaining \eqref{Peregrine_2+1},\eqref{Peregrine_3+1}. The study of the first few terms of the Q1D asymptotic expansion having \eqref{Peregrine_2+1} as leading order term is presented in \cite{CS6}.   
\vskip3pt
\noindent
\textbf{Genericity of fission}. AW fission is by no means restricted to the Q1D regime. Indeed the MI domain in Fourier space of the MNLS equation is naturally divided into two finite subdomains, and AW fission occurs in the internal one, containing the Q1D narrow strip around the segment $|k_x|<2,k_y=0$. In the external one, AW fission does not occurs, and the NLSMI is characterized, as in $1+1$ dimensions, by an AW growth from the background, followed by its decay to the background. See Figure \ref{HNLS_fission_region}) and \cite{CS5} for more details.
\vskip3pt
\noindent
\textbf{AW recurrence}. The first NLSMI of Q1D AWs is essentially the same for all examples of MNLS equations, highlighting the universal nature of the Q1D regime, and this result is explained by the qualitative argument that, in the Q1D regime, all MNLS equations are close to NLS, and therefore close to each other. This argument fails when one studies the AW recurrence of $x$ - periodic Q1D AWs, that turns out to be different for different MNLS equations, and in the forthcoming paper \cite{CS7} we use the $1+1$ dimensional finite gap theory of AWs \cite{GS1,GS3}, and the $1+1$ dimensional perturbation theory of AWs \cite{CGS1} to explain quantitatively these differences.

\vskip 10pt
\noindent
{\bf Acknowledgments}. F.C. and P.M.S. have been supported by the Research Project of National Interest PRIN2020 No. 2020X4T57A; F.C. was also supported by PRIN2022 No. 20223T577Z. We also acknowledge the useful support of the INFN Research Project MMNLP.

\end{document}